\documentclass[12pt]{article}
\usepackage{amstex,amssymb,epsfig,cite}

\newcommand{\M}{{\cal M}}

\begin{document}
\begin{flushright}
SU-4240-712 \\
imsc 99/10/36 \\
\end{flushright}

\centerline{\bf The Fermion Doubling Problem and 
Noncommutative Geometry }
\bigskip
\begin{center}
A. P. Balachandran$^{*}$, T.R.Govindarajan$^{+}$, B. Ydri$^{*}$ .\\
$^{*}${\it Physics Department, Syracuse University, \\
Syracuse,N.Y.,13244-1130, U.S.A.}\\
\bigskip
$^{+}${\it Institute of Mathematical Sciences,\\
Chennai 600 113, India}\\ 
\end{center}
\vskip.5cm
\begin{abstract}
We propose a resolution for the fermion doubling problem in discrete 
field theories based on the fuzzy sphere and its Cartesian products. 

\end{abstract}

\bigskip
The nonperturbative formulation of chiral gauge theories is  a long standing 
programme in particle physics. 
It seems clear that one should regularise
these theories with all symmetries intact.
There are two major problems associated with conventional lattice
approaches  to this programme, both with roots in topological features:
(1) The well-known Nielsen-Ninomiya
theorem \cite{NN} states that if we want to maintain chiral symmetry, then under
plausible assumptions, one cannot avoid the doubling of fermions in the 
usual lattice formulations . (2) It is not straightforward to
understand features like anomalies with naive  ultraviolet cut-off.  

Recently a novel approach to discrete physics has been developed. 
It works with quantum fields
on a ``fuzzy space'' ${\M}_{F}$ obtained by treating the underlying 
manifold ${\M}$ as a phase space and quantising it
\cite{madore,gropre,grklpr1,grklpr2,grklpr3,watamura1,watamura2,frgrre}.
The earliest contributions to topological features of the emergent fuzzy physics
came from Grosse, Klim\v{c}\'{\i}k and Pre\v{s}najder \cite{grklpr1}. They dealt
with monopoles and chiral anomaly for the fuzzy two-sphere $S_F^2$ and
took particular advantage of supersymmetry .
Later  Baez et al.
\cite{bbiv} further elaborated on their monopole work and developed the  
fuzzy physics of ${\sigma}-$ models  and their solitons
using cyclic cohomology \cite{connes,coquereaux}.
An attractive feature of this cohomological approach is its
ability to write analogues of continuum winding number formulae and
derive a fuzzy Belavin-Polyakov bound \cite{bbiv}.  
 
In this paper, we propose a solution of the fermion doubling 
problem for ${\cal M}=S^2$ using fuzzy physics . An alternative approach, 
also using fuzzy physics, can be found in \cite{grklpr1} while further 
developments of our method with applications to instantons and chiral 
anomaly is reported in \cite{bal}.There have also been important 
developments \cite{luscher} in the theory of chiral fermions and 
anomalies in the usual lattice formulations. They avoid fermion 
doubling by relaxing the requirement that chirality and Dirac operators anticommute. 
In contrast, fuzzy physics needs no such relaxation. 

Quantisable adjoint orbits of compact semi-simple Lie groups seem 
amenable to the full fuzzy treatment and lead to manageable finite 
dimensional matrix models for quantum fields . There are two such 
manifolds in dimension four, namely $S^2{\times}S^2$ and 
$ {\mathbb C}{\mathbb P}^2 $ . Our methods readily extend to $S^2{\times}S^2$. 
They are not anticipated to encounter obstructions for 
$ {\mathbb C}{\mathbb P}^2 $ as well. But we have not yet fully worked out 
its noncommutative geometry. The published work of Grosse and 
Strohmaier \cite{grostr} on $ {\mathbb C}{\mathbb P}^2 $ gives their 
description of fuzzy $4d$ fermions.

A sphere $S^2$ is a submanifold of ${\mathbb R}^3$:
\begin{equation} 
S^2=\langle \vec{n} \in {\mathbb R}^3: \sum_{i=1}^3 n_i^2=1 \rangle. 
\end{equation}
If $\hat{n}_i$ are the coordinate functions on $S^2$,
$\hat{n}_i(\vec{n}) = n_i$, then $\hat{n}_i$ commute and the algebra
${\cal A}$ of smooth functions they generate is commutative.
In contrast, the operators $x_i$ describing $S_F^2$ are
noncommutative:
\begin{equation} 
[x_i, x_j] = \frac{i \epsilon_{ijk} x_k}{[l(l+1)]^{1/2}}, \quad
\sum_{i=1}^3 x_i^2 = {\bf 1}, \quad l \in \{\frac{1}{2}, 1,
\frac{3}{2} \ldots \}.
\end{equation}
The $x_i$ commute and become $\hat{n}_i$ in the limit 
$l \rightarrow \infty$. If $L_i =
[l(l+1)]^{1/2}x_i$, then $[L_i, L_j] = i \epsilon_{ijk}L_k$ and 
$\sum L_i^2 = l(l+1)$, so that $L_i$ give 
the irreducible representation (IRR)
of the $SU(2)$ Lie algebra for angular momentum $l$. $L_i$ or $x_i$
generate the algebra $A=M_{2l+1}$ of $(2l+1) \times (2l+1)$ matrices.

Scalar wave functions on $S^2$ come from elements of ${\cal A}$. In a
similar way, elements of $A$ assume the role of scalar wave functions
on $S_F^2$. A scalar product on $A$ is $\langle \xi, \eta \rangle = Tr
{\xi}^{\dagger} \eta$. $A$ acts on this Hilbert space by left- and
right- multiplications giving rise to the left and right- regular
representations $A^{L,R}$ of $A$. For each $a \in A$, we thus have
operators $a^{L, R} \in A^{L,R}$ acting on $\xi \in A$ according to
$a^L \xi = a \xi, a^R \xi = \xi a$. [Note that $a^L b^L = (ab)^L $ while $a^R
b^R = (ba)^R$.] We assume by convention that elements of $A^L$ are to
be identified with fuzzy versions of functions on $S^2$.
Note that there are two kinds of angular momentum operators 
$L_i^{L}$ and $-L_i^{R} $ for $S_F^2$. The orbital angular momentum operator, 
which should annihilate ${\bf 1}$,
is ${\cal L}_i = L_i^L - L_i^R$. $\vec{\cal L}$ plays the role of the
continuum $-i(\vec{r} \times \vec{\nabla})$.

The construction of the Dirac operator is of crucial importance for fuzzy 
physics. 
The following two Dirac operators on $S^2$ have occurred in the fuzzy literature :
\begin{eqnarray} 
{\cal D}_1 &=& \vec{\sigma}. [-i(\vec{r} \times \vec{\nabla})] + {\bf
1}, \\
{\cal D}_2 &=& \epsilon_{ijk}\sigma_i \hat{n}_j {\cal J}_k,
\end{eqnarray}
where 
\begin{equation} 
{\cal J}_k = [-i(\vec{x} \times \vec{\nabla})]_k + \sigma_k/2 =
\rm{Total~ angular~ momentum~ operators~} .
\end{equation}
There is a common chirality operator $\Gamma$ anticommuting with both:
\begin{eqnarray}	             
\Gamma = \vec{\sigma}.\hat{n} &=& \Gamma^{\dagger}, \quad \Gamma^2 =
{\bf 1},\label{g2} \\
\Gamma {\cal D}_\alpha + {\cal D}_\alpha \Gamma &=&0.
\end{eqnarray} 
These Dirac operators {\it in the continuum} are unitarily
equivalent,
\begin{eqnarray}
{\cal D}_2& =& \exp{(i \pi \Gamma/4)} {\cal D}_1 \exp{(-i \pi \Gamma/4)}
\label{gamma}\nonumber\\
&=&i{\Gamma}{\cal D}_1,\nonumber\\
\end{eqnarray}
and have the spectrum 
\begin{equation}
\text{Spectrum of}\,\,{\cal D}_\alpha =  \{ \pm (j+1/2): j \in \{1/2,
3/2, \ldots \} \}, 
\end{equation}
where $j$ is total angular momentum (spectrum of $\vec{\cal J}^2 =\{j(j+1)\}$ ). 

Since $|{\cal D}_{\alpha}|$ $(\equiv $ positive square root of 
${\cal D}_{\alpha}^2$ $)$ for both ${\alpha}$ share the same spectrum and 
eigenvectors by $(9)$ and rotational invariance, $ |{\cal D}_{1}|=|{\cal D}_{2}| $. 
Further being multiples of unity for each fixed $j$, they commute 
with the rotationally invariant ${\Gamma}$. 
As they are invertible too, we have the important identity 

\begin{equation}
\Gamma~=~i{\frac{{\cal D}_1}{|{\cal D}_1|}}{\frac{{\cal D}_2}
{|{\cal D}_2|}}.
\end{equation}

The discrete version of ${\cal D}_1$ is: 

\begin{equation} 
D_1 = \vec{\sigma}. \vec{\cal L} + {\bf 1}.
\end{equation} 
Its spectrum is 
 
\begin{eqnarray}
\text{Spectrum of}\,\,D_1 &=&  \left\{ \pm (j+\frac{1}{2}): j \in
                              \{ \frac{1}{2}, \frac{3}{2}, \ldots
                              2l-\frac{1}{2} \} \right\} \nonumber \\  
                        &\cup& \left\{ (j+\frac{1}{2}):
                              j=2l+\frac{1}{2} \right\}.\label{specd1}
\end{eqnarray}
It is easy to derive (\ref{specd1}) by writing 

\begin{eqnarray}
D_1&=& \vec{J}^2 -
\vec{\cal L}^2 - \left(\frac{\vec{\sigma}}{2}\right)^2 + {\bf 1}, \\
\left(\frac{\vec{\sigma}}{2}\right)^2 &=& \frac{3}{4} {\bf 1},\\
{\cal L}_k +\frac{{\sigma}_k}{2}&=&J_k = ~\text{Total} ~\text{angular} ~\text{momentum} ~\text{operators}. 
\end{eqnarray}
 We let $j(j+1)$ denote the eigenvalues of ${\vec{J}}^2$.
Then for $\vec{\cal L}^2 = k(k+1), k \in \{0, 1, \ldots 2l \}$, if
$j=k+1/2$ we get $+(j+1/2)$ as eigenvalue of $D_1$, while if $j=k-1/2$ we get
$-(j+1/2)$. The absence of $-(2l+1/2)$ in (\ref{specd1}) is just
because $k$ cuts off at $2l$.(The same derivation works also for ${\cal D}_1$).

The discrete version of ${\cal D}_2$ is :

\begin{equation}
D_2 = -\epsilon_{ijk}\sigma_i x^L_j J_k = \epsilon_{ijk}\sigma_i 
        x_j^L L_k^R .
\end{equation}

$D_2$ is  no longer unitarily equivalent to $D_1$, its spectrum  
being \cite{watamura1,watamura2}
\begin{eqnarray}
\text{Spectrum of}\,\,D_2 &=&  \left\{ \pm(j+1/2)\left[1 +
                              \frac{1-(j+1/2)^2}{4l(l+1)}\right]^{1/2}:
                              \nonumber \right.\\
&j& \left. \in \{ \frac{1}{2},\frac{3}{2}, \ldots 2l+\frac{1}{2} \}
                              \right\}{\cup}\{0:j=2l+\frac{1}{2}\}. \label{specd2}
\end{eqnarray}

The first operator has been used extensively by Grosse et al
\cite{grklpr1,grklpr2,grklpr3} while the second was first introduced by Watamuras
\cite{watamura1,watamura2}.
It is remarkable that the eigenvalues (\ref{specd1}) coincide {\it exactly} with
those of ${\cal D}_\alpha$ upto $j=(2l-1/2)$. 
In contrast $D_2$  
has zero modes when $j~=~2l~+\frac{1}{2}$ and very small eigenvalues 
for large values of $j$, both being absent for ${\cal D}_{\alpha}$ .  
So $D_1$ is a better approximation to ${\cal D}_\alpha$.

But $D_1$ as it stands admits no chirality operator
anti-commuting with it. This is easy to
see as its top eigenvalue does not have its negative
in the spectrum. Instead  $D_2$ has the nice 
feature of admitting a chirality operator: the eigenvalue
for top $j$, even though it has no pair, is  exactly zero. 
So the best fuzzy Dirac operator has to combine the good
properties of $D_1$ and $D_2$. 
We suggest it to be $D_1$ after projecting out its top
$j$ mode. We will show that it then admits a chirality
with the correct continuum limit .

The chirality
operator anticommuting with $D_2$ and squaring to ${\bf 1}$
in the {\it entire} Hilbert space is
\begin{eqnarray}
\gamma_2 &=& \gamma_2^{\dagger} = -\frac{\vec{\sigma}.{\vec{L}}^R
-1/2}{l+1/2}, \\
\gamma_2^2 &=& {\bf 1}.
\end{eqnarray}
An interpretation of ${\gamma}_2$ is that $(1{\pm}{\gamma}_2)/2$ are projectors to subspaces where $(-\vec{L}^R+\vec{\sigma}/2)^2$ have values $(l{\pm}\frac{1}{2})(l{\pm}\frac{1}{2}+1)$ \cite{bbiv}.
From $(16)$ follows the identity

\begin{equation}
[D_1,\gamma_2]~=-~2~i~{\lambda}~D_2 ; {\lambda}~=~\sqrt{1-\frac{1}{(2l~+~1)^2}}~.
\end{equation}
Now $D_{\alpha}^2$ and $ |D_{\alpha}|(\equiv $ nonnegative square root of $D_{\alpha}^2)$ are multiples of identity for each fixed {\it j}, and ${\gamma}_2$ commutes with ${\vec{J}}$ . Hence they mutually commute :

\begin{equation}
[A,B]~=~0~ \text{for}~ A~,B~=~D_{\alpha}^2~ , |D_{\alpha}|~ \text{or}~ {\gamma}_2~ .
\end{equation}
Therefore from $(20)$,

\begin{equation}
\{D_1,D_2\}=\frac{1}{2i{\lambda}}[D_1^2,{\gamma}_2]=0 .
\end{equation}

In addition we can see that 

\begin{equation}
[D_{\alpha}^2,D_{\beta}]=[|D_{\alpha}|,D_{\beta}]=0.
\end{equation} 
If

\begin{eqnarray}
\epsilon_\alpha &=& \frac{D_\alpha}{|D_\alpha|}
                                             \quad \text{on the subspace $V$ with}\;j \leq 2l-1/2 ,
                                              \nonumber \\ 
 &=& 0 \quad \text{on the subspace $W$ with}\;j=2l+1/2.
\end{eqnarray}
it follows that 

\begin{equation}
e_1={\epsilon}_1~ , e_2={\epsilon}_2~ , e_3=i{\epsilon}_1{\epsilon}_2 
\end{equation}
generate a Clifford algebra on $V$ . That is, if $P$ is the orthogonal projector on $V$,

\begin{eqnarray} 
P \xi &=& \xi, \quad \xi \in V, \nonumber \\
      &=& 0,  \quad \xi \in W,
\end{eqnarray}
then

\begin{equation}
\{e_{\alpha},e_{\beta}\}=2{\delta}_{{\alpha}{\beta}}P.
\end{equation}

All this allows us to infer that $\{e_3,D_1\}=0$ so that it is a chirality operator for either $D_1$ or its restriction $PD_1P$ to $V$.In addition,{\it in view of $(10,25)$, it  has the correct continuum limit as well }so that it is a good choice for chirality in that respect too .

A unitary transformation of $e_3$ and $D_1$ will not disturb their nice features. Such a transformation bringing $e_3$ to ${\gamma}_2$ on $V$ is convenient . It can be constructed as follows . $e_{\alpha}$ and ${\gamma}_2$ being rotational scalars leave the two-dimensional subspaces in $V$ with fixed values  of $\vec{J}^2$ and $J_3$ invariant . On this subspace, $e_{\alpha}$ and unity form a basis for linear operators, so ${\gamma}_2$  is their linear combination . As $e_1$,$e_3$ and ${\gamma}_2$ anticommute with $e_2$,and all square to ${\bf 1}$, in this subspace , we infer that ${\gamma}_2$ is a transform by a unitary operator $U=exp(i{\theta}e_2/2)$ of $e_3$ in each such subspace . And ${\theta}$ can depend only on $\vec{J}^2$ by rotational invariance . Thus we can replace $PD_1P$ and $e_3$ by the new Dirac and chirality operators

\begin{eqnarray} 
D &=& e^{(i \theta (J^2) \epsilon_2)/2} (P D_1 P) e^{(-i \theta (J^2)
\epsilon_2)/2},\nonumber\\
{\gamma}:&=&P\gamma_2 P = e^{(i \theta (J^2) \epsilon_2)/2}
(i\epsilon_1\epsilon_2) e^{(-i \theta (J^2) \epsilon_2)/2} \nonumber \\
   &=& \cos \theta (J^2) (i\epsilon_1\epsilon_2) + \sin \theta (J^2)
\epsilon_1. 
\end{eqnarray}
The coefficients can be determined by taking traces with ${\epsilon}_1$ and $i{\epsilon}_1{\epsilon}_2$ .

We have established that chiral fermions can be defined on $S^2_F$ with no 
fermion doubling, at least in the absence of fuzzy monopoles.
 We next extend this result to include them as well.

\paragraph{\it{Monopoles and Instantons}:}
 
In the continuum, monopoles and instantons are particular connection fields $\omega$
on certain twisted bundles over the base manifold ${\cal M}$. On
$S^2$, they are monopole bundles, on $S^4$ or ${\mathbb C}{\mathbb
P}^2$, they can be $SU(2)$ bundles. 

In algebraic $K$-theory, it is
well-known that these bundles are associated with projectors ${\cal
P}$ \cite{connes,coquereaux,mssv}.  ${\cal P}$ is a matrix of some dimension $M$ with ${\cal P}_{ij}
\in {\cal A} \equiv {\cal C}^{\infty}({\cal M})$, ${\cal P}^2 = {\cal
P} = {\cal P}^{\dagger}$. The physical meaning of ${\cal P}$ is the
following. Let ${\cal A}^M = {\cal A} \otimes {\mathbb C}^M = \{ \xi
=(\xi_1, \xi_2 \ldots \xi_M):\xi_i \in {\cal A} \}$. Then ${\cal PA}^M
= \{ {\cal P}\xi: \xi \in {\cal A}^M\}$ consists of smooth sections
(or wave functions) ${\cal P} \xi$ of a vector bundle over ${\cal
M}$. For suitable choices of ${\cal P}$, we get monopole or instanton
vector bundles. These projectors are known \cite{mssv} and
those for monopoles have been reproduced in \cite{bbiv}.

The projectors $p^{(\pm N)}$ for fuzzy monopoles of charge $\pm N$
have also been explicitly found in \cite{bbiv}. They act on $A^{2^N} = \{ \xi$
with components $\xi_{b_1 \ldots b_N} \in A, b_i \in \{1,2\} \}$.
Let $\vec{\tau}^{(i)}$ ($i=1, 2, \ldots N$) be commuting Pauli
matrices. $\vec{\tau}^{(i)}$ has the normal action on the index $b_i$
and does not affect $b_j$ ($j \neq i$). Then $\vec{K} = \vec{L^L} +
\sum_i \vec{\tau}^{(i)}/2$ generates the $SU(2)$ Lie algebra and
$p^{(N)}$ ($p^{(-N)}$) is the projector to the maximum (minimum)
angular momentum $k_{\text{max}}=l+N/2$
($k_{\text{min}}=l-N/2$). [$\vec{K}^2 p^{(N)} =
k_{\text{max}}(k_{\text{max}}+1) p^{(N)}$, $\vec{K}^2 p^{(-N)} =
k_{\text{min}}(k_{\text{min}}+1) p^{(-N)}$.] Fuzzy analogues of
monopole wave functions are $p^{(\pm N)} A^{2^N}$. 
When spin is included, we must enhance $p^{(\pm N)} A^{2^N}$ to
$p^{(\pm N)} A^{2^N} \otimes {\mathbb C}^2 = p^{(\pm N)} A^{2^{N+1}} =
\{\xi$ with components $\xi_{b_1 \ldots b_N, j} \in A: b_i, j \in
\{1,2 \} \}$.

The complications to be resolved now are caused by the need to project
out a subspace of $A^{2^{N+1}}$. It is the analogue of the subspace
projected out by $P$ for $N=0$. In its absence, for example in the
continuum, there is a canonical way due to Connes 
\cite{connes} for extending  cyclic cohomology to
twisted bundles.

In the $N=0$ sector, the projector $P$ cuts out the 
subspace $W$ of $A^2$. 
When we pass to $p^{(\pm N)} A^{2^N}$ and thence to $p^{(\pm N)}
A^{2^{N+1}}$ by including spin, the subspace to be projected out is
{\it not} determined by $P$ if $N \neq 0$, as we shall see
below. Rather, we can explain it as follows: Let $\vec{J} = \vec{K} -
\vec{L^R} + \vec{\sigma}/2$ be the ``total angular momentum''. Calling
$\vec{J}$ by this name is appropriate as its components become $(15)$ for
$N=0$ and displays the known ``spin-isospin mixing''\cite{jacreb}
for $N \neq 0$. The maximum of $\vec{J}^2$ on
$p^{(\pm N)} A^{2^{N+1}}$ is $J_{\text{max}}(J_{\text{max}} +1),
J_{\text{max}} = (l \pm N/2)+l+1/2 =2l \pm N/2 +1/2$. [We assume that
$2l \geq (N-1)/2$.] The vectors to be projected out are those with
total angular momentum $J_{\text{max}}$. If ${\cal J}^{(\pm N)}$ are
the corresponding projectors [with ${\cal J}^{(0)} = P$], the twisted 
space we work with is ${\cal J}^{(\pm N)}p^{(\pm N)}
A^{2^{N+1}}$. Since $p^{(\pm N)}$ commute with $\vec{J}$ and hence with
${\cal J}^{(\pm N)}$, $Q^{(\pm N)} = {\cal J}^{(\pm N)}p^{(\pm
N)}$ are also projectors. 

There is no degeneracy for angular momentum $J_{\text{max}}$ in
$p^{(\pm N)} A^{2^{N+1}}$. That is because there is only one way to
couple $l \pm N/2, l$ and $1/2$ to $J_{\text{max}}$. The space $({\bf
1} - {\cal J}^{(\pm N)})p^{(\pm N)} A^{2^{N+1}}$ is thus of dimension
$2 J_{\text{max}} + 1$. We want to get rid of this subspace.

The operators $T = D$ or $\gamma$ are zero on $({\bf
1} - P)A^2$ where $P$ cuts out states of angular momentum
$2l+1/2$. There is no degeneracy for this angular momentum in
$A^2$. $T$ and $P$ extend canonically to $A^2 \otimes {\mathbb
C}^{2^N} (\equiv A^{2^{N+1}})$ as $T \otimes {\bf 1}$ and $P \otimes
{\bf 1}$. Let us call them once more as $T$ and $P$. $T$ and $P$
commute with $\vec{J}$ and hence with ${\cal J}^{(\pm N)}$. There is
only one way to couple $N$ ``isospin'' $1/2$'s to $(2l + 1/2)$ to get
$J_{\text{max}}$ so that $({\bf 1} - {\cal J}^{(\pm N)})({\bf 1} - P)
A^{2^{N+1}}$ is also of dimension $2 J_{\text{max}}+1$. And $T$ is
zero on this subspace.

The projectors $({\bf 1} - {\cal J}^{(\pm N)})p^{(\pm N)}$ and $({\bf
1} - {\cal J}^{(\pm N)})({\bf 1} - P)$ being of the same rank, there
exists a unitary operator $U$ on $A^{2^{N+1}}$ transforming one to the
other:
\begin{equation} 
({\bf 1} - {\cal J}^{(\pm N)})p^{(\pm N)} = U({\bf 1} - {\cal J}^{(\pm
N)})({\bf 1} - P)U^{-1}. 
\label{Udef}
\end{equation} 
If we transport $T$ by $U$,
\begin{equation} 
T' = UTU^{-1},
\end{equation} 
then $T'=D', F'$ or $\gamma'$ vanishes on $({\bf 1} - {\cal J}^{(\pm
N)}) p^{(\pm N)} A^{2^{N+1}}$. On its orthogonal complement $[{\bf 1}
- ({\bf 1} - {\cal J}^{(\pm N)})p^{(\pm N)}] A^{2^{N+1}}$, invariant
under $T'$, $\gamma'$ squares to unity and 
anticommutes with $D'$ just as we want. What replaces
$P$ now is not ${\cal J}^{(\pm N)}$, but rather
\begin{eqnarray} 
P^{(\pm N)} &=& [{\bf 1}-({\bf 1}-{\cal J}^{(\pm N)})p^{(\pm N)}], \\ 
P^{(0)} &=& P.
\end{eqnarray} 

As $l \rightarrow \infty$, $J_{\text{max}}$ becomes dominated by $2l$
and so we have the freedom to let $U$ approach ${\bf 1}$. That is, no
$U$ is needed in the continuum limit.

Total angular momentum $2l+1/2+N/2$ has no multiplicity in
$A^{2^{N+1}}$. As both $({\bf 1} - {\cal J}^{(N)})({\bf 1} - P)
A^{2^{N+1}}$ and $({\bf 1} - {\cal J}^{(N)})p^{(N)} A^{2^{N+1}}$ have
this angular momentum, we have that 
\begin{equation} 
({\bf 1} - {\cal J}^{(N)})({\bf 1} - P) A^{2^{N+1}} = 
({\bf 1} - {\cal J}^{(N)})p^{(N)} A^{2^{N+1}}.
\label{same}
\end{equation}
So we choose
\begin{equation} 
U={\bf 1} \quad \text{on} \quad ({\bf 1} - {\cal J}^{(N)})({\bf 1}-P)
A^{2^{N+1}}. 
\label{Uchoice}
\end{equation}

Next in accordance with (\ref{Udef}), we set 
\begin{equation} 
U({\bf 1} - {\cal J}^{(-N)})({\bf 1} - P) A^{2^{N+1}} =
({\bf 1} - {\cal J}^{(-N)})p^{(-N)} A^{2^{N+1}}.
\label{UonminusN}
\end{equation}
We also demand that 
\begin{equation}
[U, J_i]=0.
\label{rotinv}
\end{equation}
That fixes $U$ upto a phase on the subspace $({\bf 1} - {\cal
J}^{(-N)})({\bf 1} - P) A^{2^{N+1}}$.

We saw in \cite{bbiv} that $(1 \pm \gamma)/2$ are projectors for
combining $-\vec{L}^R$ and $\vec{\sigma}/2$ to give angular momenta $l
\pm 1/2$. So $\gamma =+1$ on all the subspaces $({\bf 1} - {\cal
J}^{(\pm N)})({\bf 1}-P) A^{2^{N+1}}$, $({\bf 1} - {\cal J}^{(\pm
N)})(p^{(\pm N)}) A^{2^{N+1}}$. Also, $(\frac{\sum
\vec{\tau}^{(i)}}{2})^2$ is $\frac{N}{2}(\frac{N}{2}+1)$ on these
subspaces. It follows that (\ref{Uchoice}), (\ref{UonminusN}) and
(\ref{rotinv}) are compatible with a $U$ commuting with $J_i$,
$\gamma$, $(-\vec{L}^R +\vec{\sigma}/2)^2$ and $(\frac{\sum
\vec{\tau}^{(i)}}{2})^2$. We now outline an extension of $U$ to all of
$A^{2^{N+1}}$ consistently with rotational invariance (\ref{rotinv})
and
\begin{equation} 
\left[U, \left( -\vec{L}^R+\frac{\vec{\sigma}}{2} \right)^2 \right] = 
\left[U, \left(\frac{\sum \vec{\tau}^{(i)}}{2}\right)^2 \right]= 0.
\end{equation}
An important consequence is that
\begin{equation} 
\gamma' = \gamma
\end{equation}
so that 
\begin{equation}
[\gamma', p^{(\pm N)}] = [\gamma', J_i] = [\gamma', {\cal J}^{(\pm
N)}] = 0.
\label{chiprop}
\end{equation}

One way to specify $U$ more fully is as follows. Let 
\begin{equation} 
A^{2^{N+1}} = X \oplus X^{\perp} = X' \oplus {X'}^{\perp}
\end{equation}
be orthogonal decompositions where 
\begin{eqnarray}	
X &=& ({\bf 1}-{\cal J}^{(N)})({\bf 1} - P)A^{2^{N+1}} \oplus ({\bf
1}-{\cal J}^{(-N)})({\bf 1} - P)A^{2^{N+1}}, \\
X' &=& UX = ({\bf 1}-{\cal J}^{(N)})p^{(N)}A^{2^{N+1}} \oplus ({\bf
1}-{\cal J}^{(-N)}) p^{(-N)} A^{2^{N+1}}.
\end{eqnarray}
Both $X$ and $X'$ are invariant under the self-adjoint operators
\begin{equation} 
J_i, \left( -\vec{L}^R + \frac{\vec{\sigma}}{2} \right)^2 \quad 
\text{and} \quad \left( \sum \frac{\vec{\tau}^{(i)}}{2} \right)^2. 
\nonumber 
\end{equation}
Therefore, the same is the case with $X^{\perp}$ and
${X'}^{\perp}$. We can extend $U$ to a map $X^{\perp} \rightarrow
{X'}^{\perp}$ which commutes with the above operators. There would
still be uncertainties about choosing $U$ requiring further
conventions for elimination.

Although $T'$ are operators on $P^{(\pm N)}A^{2^{N+1}}$,
that is not the space of sections for the twisted bundles. The latter
is, rather, 
\begin{equation}
Q^{(\pm N)} A^{2^{N+1}} =p^{(\pm N)}P^{(\pm N)} A^{2^{N+1}}= p^{(\pm N)}{\cal J}^{(\pm N)}A^{2^{N+1}} .
\end{equation} 
It is not an invariant subspace for $D'$ unless $D'$ is 
projected, or corrected by connections as explained in \cite{bal} . However, chirality $\gamma'$ is well-defined on twisted sections
because of (\ref{chiprop}).

We now permanently rename $T'$, $P^{(\pm
N)}$ and $Q^{(\pm N)}$ as follows:
\begin{eqnarray}	
D',  \gamma' &\rightarrow& D, \gamma~ ~,~ ~p^{(\pm N)} \rightarrow p, \nonumber\\
P^{(\pm N)} &\rightarrow& P~ ~,~ ~ Q^{(\pm N)} \rightarrow Q.
\end{eqnarray} 
$\gamma'$ in any case is $\gamma$.

This completes our construction of Dirac operator and chirality on $S^2_F$ . We may remark here that the algebra for the space $S^2_F{\times}S^2_F$ is $A{\otimes}_{\mathbb C}A$ while its Dirac and chirality operators are $D{\otimes}{\bf 1} +{\gamma}{\otimes}D$ and ${\gamma}{\otimes}{\gamma}$ .

\noindent {\bf Acknowledgments} 

S.Vaidya, Xavier
Martin, Denjoe O'Connor and Peter Pre\v{s}najder offered us many good
suggestions during this work while Apoorva Patel told us about
\cite{luscher}. We thank them for their help. The work of
APB was supported in part by the DOE under contract number
DE-FG02-85ER40231.

\bibliographystyle{unsrt}

\begin{thebibliography}{10}

\bibitem{NN}
N. B. Nielsen and M. Ninomiya, 
\newblock {\em Phys.Lett.}, B105: 219, 1981;
\newblock {\em Nucl.Phys.}, B185: 20, 1981.

\bibitem{madore}
J.~Madore.
\newblock {\em An Introduction to Noncommutative Differential Geometry
and its Applications}.
\newblock Cambridge University Press, Cambridge, 1995;
{\tt gr-qc/9906059}.

\bibitem{gropre}
H.~Grosse and P.~Pre\v{s}najder.
\newblock {\em Lett.Math. Phys.}, 33:171--182, 1995.
\newblock and references therein.

\bibitem{grklpr1}
H.~Grosse, C.~Klim\v{c}\'{\i}k, and P.~Pre\v{s}najder.
\newblock {\em Commun.Math.Phys.}, {\bf 178}:507--526, 1996; {\bf
185}:155--175, 1997;
\newblock H.~Grosse and P.~Pre\v{s}najder. 
\newblock {\em Lett.Math.Phys.} {\bf 46}:61--69, 1998 and ESI preprint,
1999.  

\bibitem{grklpr2}
H.~Grosse, C.~Klim\v{c}\'{i}k, and P.~Pre\v{s}najder.
\newblock {\em Commun.Math.Phys.}, 180:429--438, 1996.
\newblock {\tt hep-th/9602115}.
 
\bibitem{grklpr3}
H.~Grosse, C.~Klim\v{c}\'{i}k, and P.~Pre\v{s}najder.
\newblock In {\em Les Houches Summer School on Theoretical Physics}, 1995.
\newblock {\tt hep-th/9603071}.
\newblock See citations in \cite{gropre,grklpr1,grklpr2,grklpr3} for
further references.
 
\bibitem{watamura1}
U.~Carow-Watamura and S.~Watamura.
\newblock {\tt hep-th/9605003}.
\newblock {\em Commun.Math.Phys.},183:365--382, 1997.
 
\bibitem{watamura2}
U.~Carow-Watamura and S.~Watamura.
\newblock {\tt hep-th/9801195}.
 
\bibitem{frgrre}
J.~Frohlich, O.~Grandjean, and A.~Recknagel.
\newblock {\em Commun. Math. Phys.}, 203:119--184,, 1999.
\newblock {\tt math-phy/9807006} and references therein for fuzzy group
manifolds.

\bibitem{bbiv}
S.~Baez, A.~P. Balachandran, S.~Vaidya and B.~Ydri.
\newblock {\tt hep-th/9811169} and Comm.Math.Phys (in press). 

\bibitem{connes}
A.~Connes.
\newblock {\em Noncommutative Geometry}.
\newblock Academic Press, London, 1994;
G.~Landi.
\newblock {\em An Introduction to Noncommutative Spaces And Their Geometries}.
\newblock Springer-Verlag , Berlin , 1997 .
\newblock {\tt hep-th/9701078}.

 
\bibitem{coquereaux}
R.~Coquereaux.
\newblock {\em J. Geo. Phys.}, 6:425--490, 1989.


\bibitem{bal}
A. P. Balachandran, S. Vaidya.
\newblock {\em SU-4210-701 ,TIFR/TH/99-32 } and {\tt hep-th/9910129}.

\bibitem{luscher}
M. Luscher.
\newblock {\em Phys. Lett. {\bf B428} (1998) 342} and
\newblock {\tt hep-lat/9802011} ; {\tt hep-lat/9904009} and references therein .
\bibitem{grostr}
H.~Grosse and A.~Strohmaier.
\newblock {\em Lett.Math.Phys.} {\bf 48}:163--179,1999. 
\newblock {\tt hep-th/9902138}.

\bibitem{mssv}
J. A. Mignaco, C. Sigaud, A. R. da Silva and F. J. Vanhecke.
\newblock {\em Rev.Math.Physics, {\bf 9} (1997) 689}
and {\tt hep-th/9611058};
G. Landi.
\newblock {\tt math-phy/9812004; math-phy/9905014}.


\bibitem{jacreb}
R.~Jackiw and C.~Rebbi.
\newblock {\em Phys.Rev.Lett.} {\bf 36}:1116-1119, 1976 ;
P.~Hasenfratz and G. 't~Hooft.
\newblock {\em Phys.Rev.Lett.} {\bf 36}:1119-1122, 1976.

\end{thebibliography}

\end{document}